# Tuning magnetic and transport properties in quasi-2D $(Mn_{1-x}Ni_x)_2P_2S_6$ single crystals

Y. Shemerliuk[a], Y. Zhou[d] Z. Yang[d], G. Cao[c], A. U. B. Wolter[a], B. Büchner[a,b], S. Aswartham[a,*]

[a] *Institut für Festkörperforschung, Leibniz IFW Dresden, Helmholtzstraße 20, 01069 Dresden, Germany*
[b] *Institut für Festkörper- und Materialphysik and Würzburg-Dresden Cluster of Excellence ct.qmat, Technische Universität Dresden, 01062 Dresden, Germany*
[c] *University of Colorado at Boulder, Boulder, USA*
[d] *Anhui Province Key Laboratory of Condensed Matter Physics at Extreme Conditions, High Magnetic Field Laboratory, Chinese Academy of Sciences, Hefei 230031, China*

## Abstract

We report an optimized chemical vapor transport method to grow single crystals of $(Mn_{1-x}Ni_x)_2P_2S_6$ where x = 0, 0.3, 0.5, 0.7 & 1. Single crystals up to 4 mm × 3 mm × 200 μm were obtained by this method. As-grown crystals are characterized by means of scanning electron microscopy, and powder x-ray diffraction measurements. The structural characterization shows that all crystals crystallize in monoclinic symmetry with the space group *C2/m* (No. 12). We have further investigated the magnetic properties of this series of single crystals. The magnetic measurements of the all as-grown single crystals show long-range antiferromagnetic order along all crystallographic principal axes. Overall, the Néel temperature $T_N$ is non-monotonous, with increasing $Ni^{2+}$ doping the temperature of the antiferromagnetic phase transition first decreases from 80 K for pristine $Mn_2P_2S_6$ (x = 0) up to x = 0.5, and then increases again to 155 K for pure $Ni_2P_2S_6$ (x = 1). The magnetic anisotropy switches from out-of-plane to in-plane as a function of composition in $(Mn_{1-x}Ni_x)_2P_2S_6$ series. Transport studies under hydrostatic pressure on the parent compound $Mn_2P_2S_6$ evidence an insulator-metal transition at an applied critical pressure of ~22 GPa.

# 1. Introduction

In recent years, research of functional two-dimensional (2D) materials has stimulated activities, aimed at synthesis of new materials and studies of their functionalities. This class has drawn a great deal of interest and attention chiefly because of their unique electronic [1-3], magnetic [4-8] and optical properties [10-12] to the bulk counterpart. Layered van der Waals (vdW) materials dominate among the current 2D materials. The vdW interaction only weakly couples the atomic layers together along the *c* axis, even in a bulk vdW material, and as a result, the electron confinement in the 2D lattice often leads to some specific properties, such as anisotropic magnetic behavior [13,14] and an anisotropic conductivity [15-17]. Fabricating heterostructures and optoelectronic nanodevices using these materials has resulted in impressive findings in different fields [18-22].

The transition metal phosphorus trichalcogenides (*TM*$_2$P$_2$*Ch*$_6$ *TM*=transition metal; *Ch*=chalcogen) family belongs to the layered van der Waals materials class [23]. All of these materials have very similar crystal structures, a monoclinic unit cell with the space group of *C2/m* but different magnetic properties depending on the underlying 3*d* transition metal [24]. The *TM*$_2$P$_2$*Ch*$_6$ compounds are all semiconductors at ambient pressure [23], with band gaps for most of them larger than 1 eV as determined by optical measurements [25] with extremely high room-temperature resistivities.

Band structure calculations [26,27] lead to the conclusions that *TM*$_2$P$_2$*Ch*$_6$ are Mott insulators, and by applying external pressure, these materials could be driven to an insulator-metal or Mott transition [28]. These results further stimulated substantial interest to tune the parameters of the system to an intermediate state to access the physics that is not yet fully understood. Additionally, many superconducting materials are low dimensional and lie close to antiferromagnetic Mott-insulator phases in phase diagrams. Theoretical calculations [29] show that these states increase the possibility of introducing superconductivity. Thus, tunable 2D antiferromagnetic Mott insulators provide a promising playground to investigate the basic mechanisms of several unsolved problems in condensed matter physics.

Field-effect transistors based on bulk and few-layer Mn$_2$P$_2$S$_6$ have been fabricated [30]. These devices show the potential for good ultraviolet photodetectors. Bulk and few-layer Mn$_2$P$_2$S$_6$ on top of indium titanium oxide (ITO) coated Si substrate also shows tunneling transport phenomena [31]. The estimated barrier height of thin Mn$_2$P$_2$S$_6$ flakes is 1.31 eV (±0.01) [31]. These investigations bring the magnetic van der Waals material $Mn_2P_2S_6$ very close to various applications such as field effect transistors and magnetic tunnel junction fabrication.

$Mn_2P_2S_6$ and $Ni_2P_2S_6$ are well-known members of the transition metal phosphorus trichalcogenides family. They present interesting anisotropic antiferromagnetism below the Néel temperatures of 80 K for $Mn_2P_2S_6$ and 155 K for $Ni_2P_2S_6$ [32, 33], which is not well understood. The main difference between these two pristine compounds and the key ingredient in the $TM_2P_2Ch_6$ family is the underlying anisotropy that dictates the long-range magnetic order down to the monolayer, which eventually depends on the $3d$ transition metal ions [34-37]. This anisotropy of the system can be tuned systematically by several methods, for example with chemical substitution or with applied pressure. Motivated by this, here we present two different ways of tuning the ground state of $Mn_2P_2S_6$ single crystals.

In this work we report the optimized synthesis and crystal growth conditions of $(Mn_{1-x}Ni_x)_2P_2S_6$. We have investigated the crystal structure and magnetic properties of the series of single crystals. We observe two different effects on the magnetic properties namely a shift of the long-range order temperature $T_N$ and a broad anomaly at higher temperatures above $T_N$, which becomes very sharp for selected compositions. Additional transport studies under hydrostatic pressure evidence an insulator-metal transition at a critical pressure of ~ 22 GPa.

The present article is organized in the following order. In the first section, we present the optimized methods for crystal growth for our series of crystals, $(Mn_{1-x}Ni_x)_2P_2S_6$. In the later sections results of scanning electron microscopy–energy dispersive x-ray spectroscopy (SEM-EDX), x-ray diffraction (XRD) are shown. Finally, detailed magnetic and electrical transport measurements are presented.

## 2. Experimental methods and crystal growth

The composition of our crystals was determined using energy dispersive x-ray spectroscopy (EDX), with an accelerating voltage 30 kV. Scanning electron microscope was used to obtain electron microscopic images using two types of signals: the backscattered electrons (BSE) for chemical contrast and the secondary electrons (SE) for topographic contrast. X-ray powder diffraction data were collected using a STOE powder laboratory diffractometer in transmission geometry with Cu-$K_{\alpha 1}$ radiation (the wavelength (λ) is 1.540560 Å) from a curved Ge (111) single crystal monochromator and detected by a MYTHEN 1K 12.5° linear position sensitive detector manufactured by DECTRIS. An XRD pattern of a polycrystalline sample was obtained by grinding as-grown single crystals.

Temperature and field dependent magnetization studies were performed using a Quantum Design Superconducting Quantum Interference Device Vibrating Sample Magnetometer (SQUID-VSM). The measurements were performed for field-cooled (FC) conditions and for two crystallographic directions, i.e., parallel and perpendicular to the growth direction of the crystal.

High-pressure transport experiments were performed in a screw-pressure-type DAC made of CuBe alloy. A pair of anvil culets of 300 μm was used. A mixture of epoxy and fine cubic boron nitride (*c*-BN) powder was compressed firmly to insulate the electrodes from the steel gasket. A single-crystal flake with dimensions of ~120 × 45 × 10 μm$^3$ was loaded together with NaCl fine powder and ruby powder. A five-probe configuration was utilized to measure the resistance and Hall resistance of the flake, where the external magnetic field was perpendicular to the surface of the flake. The ruby fluorescence shift was used to calibrate the pressure at room temperature in all experiments [38].

Single crystals of $(Mn_{1-x}Ni_x)_2P_2S_6$ were grown by the chemical vapor transport technique. All preparation steps were performed in a glovebox, before sealing the ampule. The proper weights of the starting materials manganese (powder, Alfa Aesar, 99.98%), nickel (powder -100 mesh, Sigma Aldrich, 99.99%), red phosphorus (lumps, Alfa Aesar, 99.999%) and sulfur (pieces, Alfa Aesar, 99.999%) with a molar ratio of $(Mn_{1-x}Ni_x)$:P:S = 1:1:3 (for x = 0; 0.3; 0.5; 0.7; 1) and a total charge mass of 1g were thoroughly mixed with iodine (0.05g). The starting material was loaded in a quartz ampoule (6 mm inner diameter, 2 mm wall thickness) and then was cooled by liquid nitrogen to avoid significant losses of the transport agent during the evacuation process to a residual pressure of $10^{-8}$ bar. The ampoule was sealed under static pressure at a length of approximately 12 cm by the oxy-hydrogen flame. Then, the closed ampules were heated in a two-zone furnace. A similar approach was used by us for the crystal growth in the closely related sister compounds such as $(Fe_{1-x}Ni_x)_2P_2S_6$ and $AgCrP_2S_6.$ [39-40].

The optimized temperatures for $Ni_2P_2S_6$ [36, 39] were chosen as 750°C and 700°C for the hot and the cold zone, respectively, and for $Mn_2P_2S_6$ [40-41] as 680°C and 630°C. For the substituted $(Mn_{1-x}Ni_x)_2P_2S_6$ samples, due to the absence of published information, the following temperature profile has been optimized by us. Initially, the furnace was heated homogeneously to 300°C with 100°C/h and dwelled for 24 hours to provide a pre-reaction of P and S with the transition elements. After that, an inverse transport gradient is applied to transport particles adhering to the walls to the one side of the ampoule which is the charge region to avoid the formation of random nucleation centers. Later, the charge region was heated to 720°C in 4 h with a dwelling time of 336 h, whereas, the sink region was initially heated up to 770°C in 4 h, and dwelled at this temperature for 24 h. Later, the temperature in the growth zone was gradually reduced during one day to 670°C to slowly form the transport temperature gradient for controlling nucleation and held at this temperature for 289 h. As a result, the temperature gradient was set for vapor transport between 720°C (charge) and 670°C (sink) for 12 days. In the final stage, the charge region was cooled to the sink temperature in 1 hour before both regions were furnace-cooled to room temperature.

Thin lustrous black and green in the $Mn_2P_2S_6$ case plate-like crystals perpendicular to the *c* axis in the size of up to 4 mm × 3 mm × 200 μm were obtained (as shown in Fig. 2). All as-grown crystals show a layered morphology and they are easily exfoliated by scotch tape.

## 3. Characterization: compositional and structural analysis

All as-grown single crystals of the series exhibit the typical features of a layered van der Waals structure, which is in line with the pristine compounds (Fig. 1). As an example, crystals are shown in Fig. 1 (a-e) with clear hexagonal layers and a well-defined flat facade with 120° angles, which clearly indicates that the crystals grew along the symmetry axes. Backscattered electron (BSE) images of our samples have no change of chemical contrast on the surface of the crystals, as shown in Fig. 1 (f-h). This indicates a homogeneous elemental composition on the respective area of the crystal.

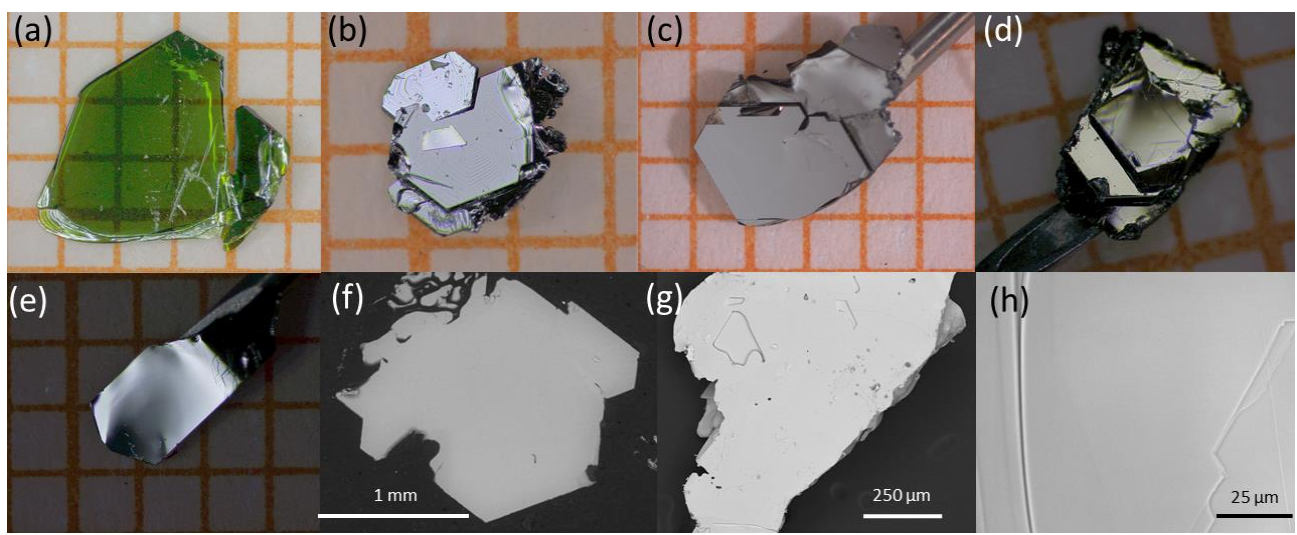

Figure 1. Examples of single crystals grown by the chemical vapor transport (a) $Mn_2P_2S_6$, (b) $(Mn_{0.7}Ni_{0.3})_2P_2S_6$, (c) $(Mn_{0.5}Ni_{0.5})_2P_2S_6$ (d) $(Mn_{0.3}Ni_{0.7})_2P_2S_6$ and (e) $Ni_2P_2S_6$; cell scale is 1 mm, SEM - images of (f)$(Mn_{0.3}Ni_{0.7})_2P_2S_6$, (g) $(Mn_{0.5}Ni_{0.5})_2P_2S_6$ and (h) $(Mn_{0.7}Ni_{0.3})_2P_2S_6$ SEM–backscattered electron (BSE) mode.

To gain further information about the distribution of elements for our $(Mn_{1-x}Ni_x)_2P_2S_6$ samples, an elemental mapping was performed on an area of the crystal with dimensions of about $150\times100\mu m^2$ as shown in Fig. 2 (a - f). The results indicate a homogeneous distribution of Mn, Ni, P and S.

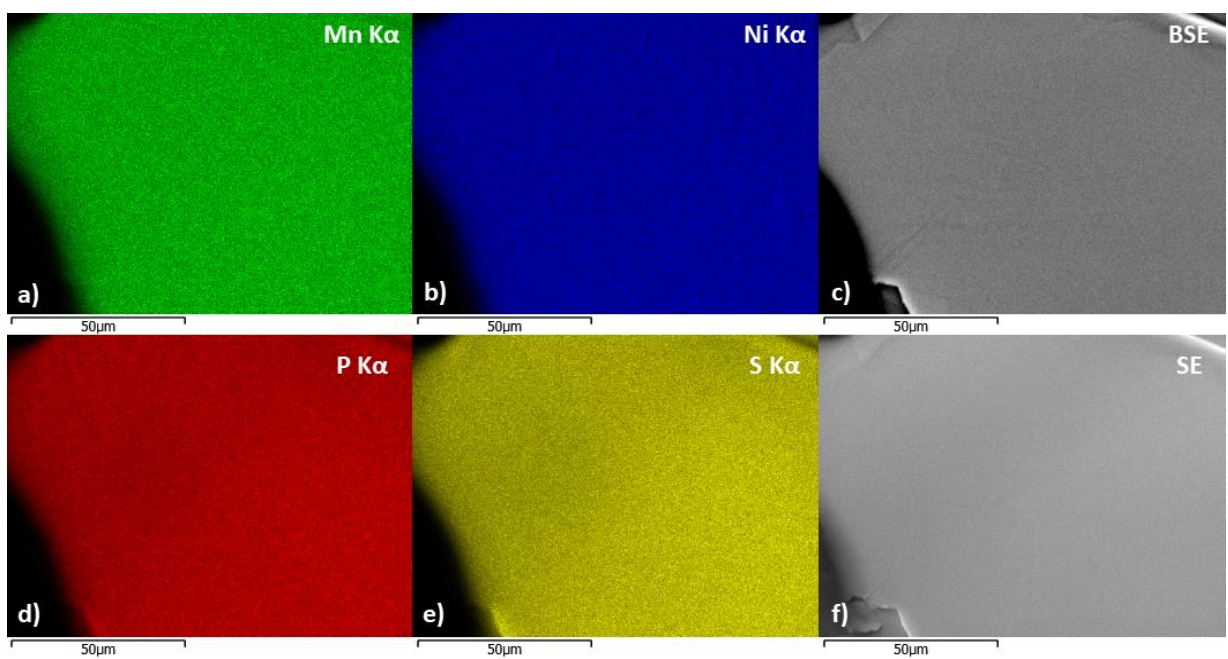

Figure 2. a), b), d), e) Elemental mapping, c) backscattered electron (BSE) image and f) secondary electron (SE) image from a $(Mn_{0.7}Ni_{0.3})_2P_2S_6$ crystal.

The composition of the as-grown single crystals was determined by energy-dispersive x-ray spectroscopy (EDX). The result of EDX measurements highly depends on the sample topography [42]. EDX was mapped out via measuring different areas and points on the surface of several crystals representing all doping concentrations (see Fig. 2 for representative results on $(Mn_{0.7}Ni_{0.3})_2P_2S_6$). As shown in Tab. 1, all $(Mn_{1-x}Ni_x)_2P_2S_6$ compounds show the expected composition. The experimental value $x_{exp}$ is found in the range of the nominal value $x_{nom}$ considering a systematic uncertainty of this ratio of approximately 5%.

Table 1. Compositional analysis of as-grown single crystals of $(Mn_{1-x}Ni_x)_2P_2S_6$.

| $x_{nom}$ | Expec. Comp. | Mean Comp. (EDX) | $x_{exp}$ |
|---|---|---|---|
| 0 | $Mn_2P_2S_6$ | $Mn_{2.02(5)}P_{2.04(1)}S_{5.93(5)}$ | 0 |
| 0.3 | $Mn_{1.4}Ni_{0.6}P_2S_6$ | $Mn_{1.45(1)}Ni_{0.55(1)}P_{2.03(6)}S_{5.96(2)}$ | 0.28 |
| 0.5 | $MnNiP_2S_6$ | $Mn_{0.80(6)}Ni_{0.91(7)}P_{2.07(9)}S_{6.19(8)}$ | 0.45 |
| 0.7 | $Mn_{0.6}Ni_{1.4}P_2S_6$ | $Mn_{0.62(4)}Ni_{1.33(3)}P_{2.04(1)}S_{6.00(2)}$ | 0.67 |
| 1 | $Ni_2P_2S_6$ | $Ni_{2.04(6)}P_{2.03(2)}S_{5.92(3)}$ | 1 |

The structural characterization and phase purity were confirmed by powder x-ray diffraction using a STOE powder diffractometer. All compositions show the same monoclinic space group *C2/m* (No. 12). There were no additional impurity phases in the pXRD patterns. Figure 3 shows the pXRD patterns for the full series of samples $(Mn_{1-x}Ni_x)_2P_2S_6$. There is a systematic shift of the all Bragg peaks toward lower angles with increasing manganese content. Significant changes in relative intensities are observed due to a strongly preferred orientation of the plate-like powder crystallites. The interlayer spacing *d* is calculated from the peak position using Bragg's law $n\lambda = 2d\sin\theta$ and plotted in Figure 4 as a function of Ni substitution for $(Mn_{1-x}Ni_x)_2P_2S_6$. The interlayer spacing and, therefore, the van der Waals gap decrease linearly with $x$ due to the smaller atomic radius of Ni compared to Mn.

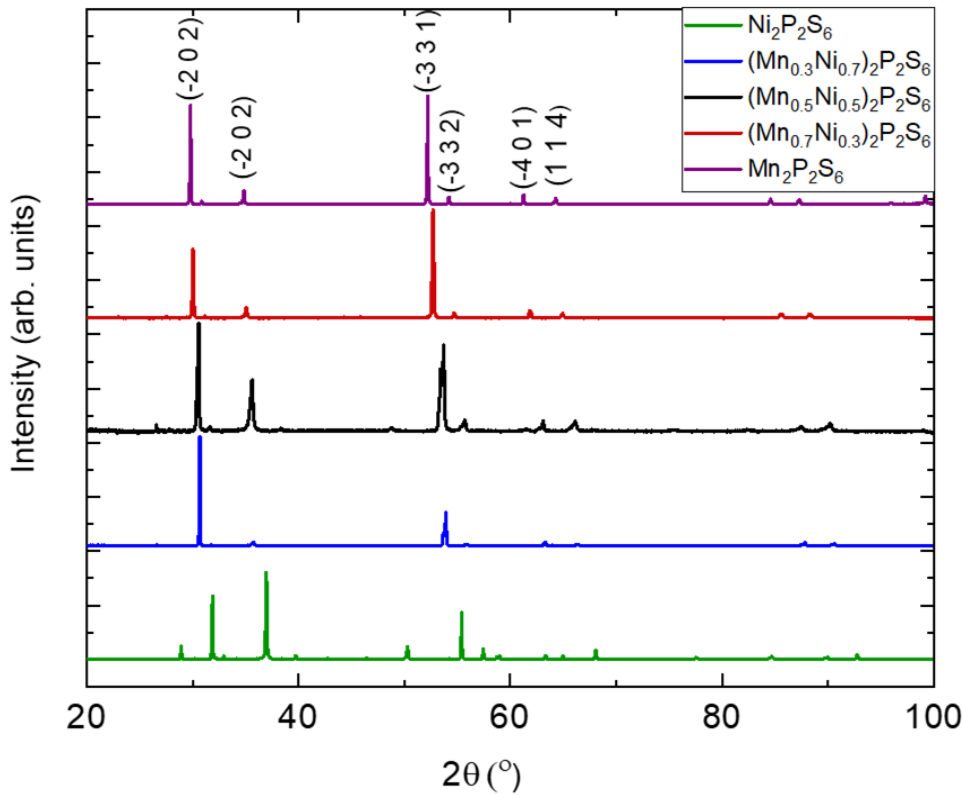

Figure 3. pXRD patterns of $(Mn_{1-x}Ni_x)_2P_2S_6$.

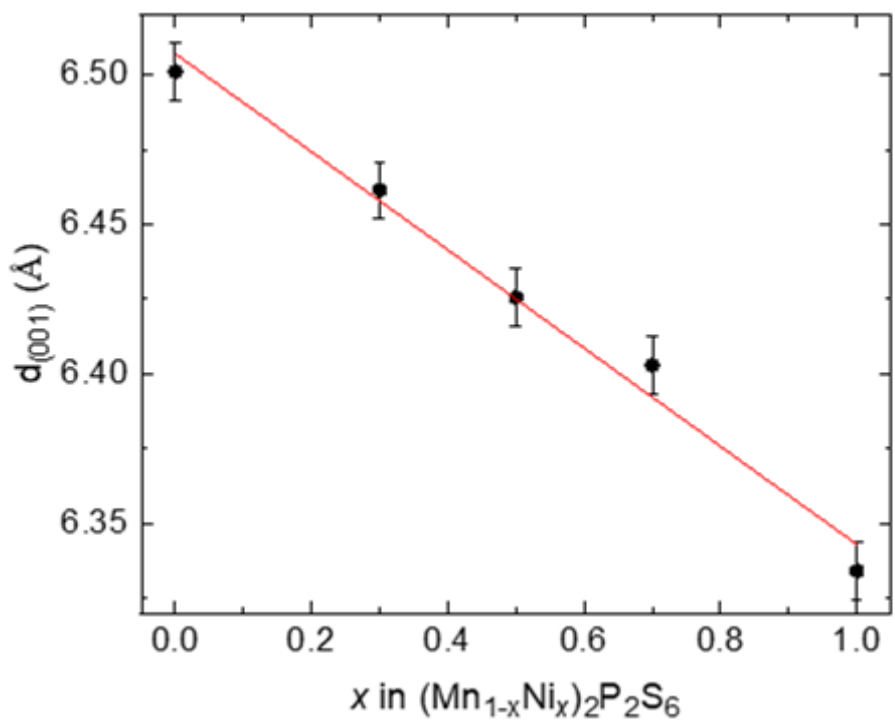

Figure 4. Interlayer spacing $d$ calculated from Bragg's law and plotted as a function of Ni substitution for $(Mn_{1-x}Ni_x)_2P_2S_6$.

All obtained pXRD patterns were described by the atomic model proposed by Klingen *et al.* for $Fe_2P_2S_6$ [43]. In this model, the transition metal ions create hexagons which are octahedrally surrounded by Sulfur. They form a honeycomb structure in the *ab* plane. P-P pairs are located in the void of each honeycomb. These layers are poorly bonded to each other by van der Waals bonds, which gives rise to a layered structure. Crystal structures projecting different crystallographic axes of $(Mn_{1-x}Ni_x)_2P_2S_6$ are shown in Fig. 5 (a-c).

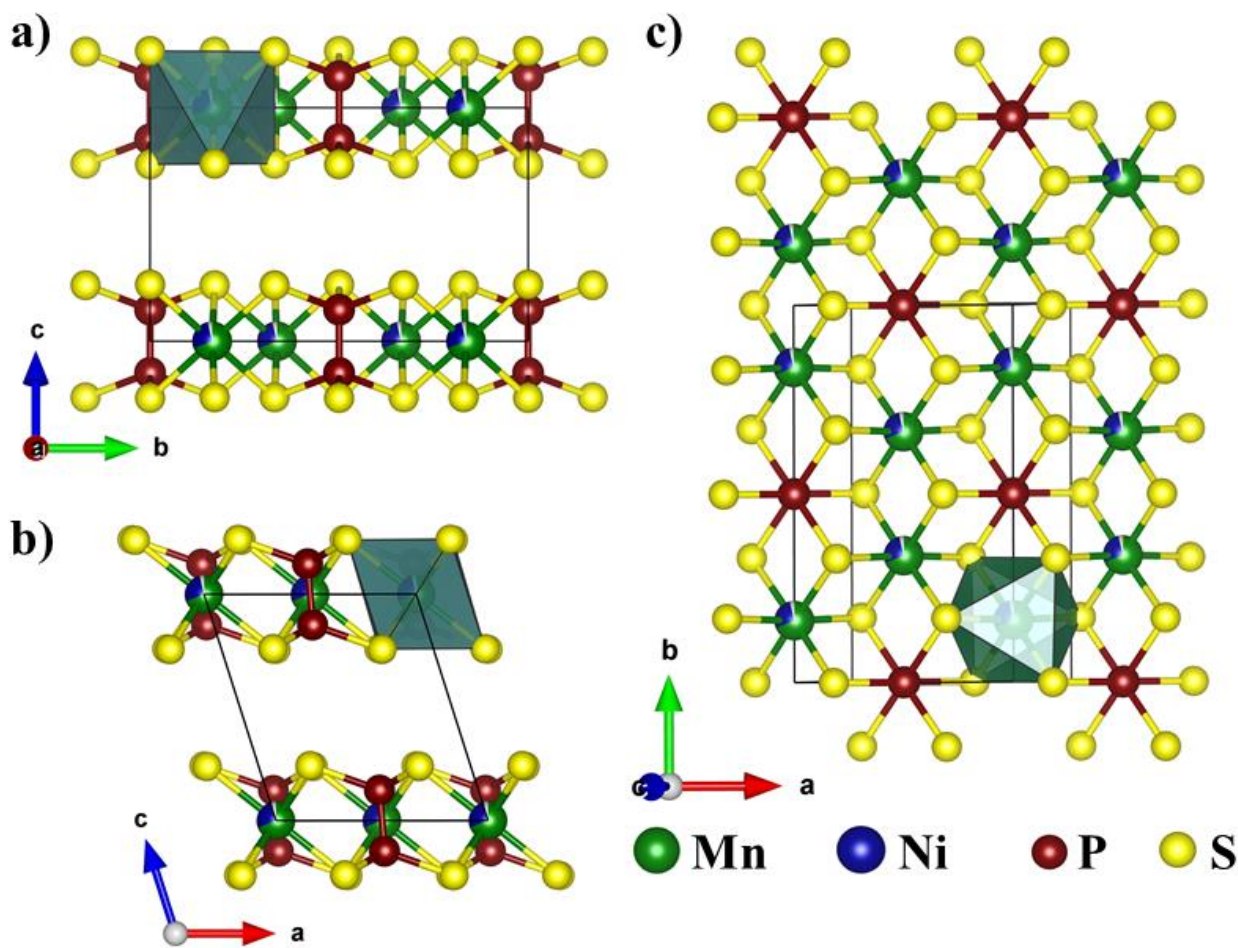

Figure 5. As an example crystal structure of $(Mn_{0.7}Ni_{0.3})_2P_2S_6$(a) shown in the *bc* plane, (b) shown in the *ac* plane and (c) shown in the *ab* plane. The graphical representation was prepared using VESTA3 [44]

The Rietveld refinement of the x-ray data was carried out using the FULLPROF software for a 30% substituted sample i.e., $(Mn_{0.7}Ni_{0.3})_2P_2S_6$. The measured pattern (red circles) is shown in Fig. 6 together with the calculated pattern based on the Rietveld analysis (black line), the difference between measured and calculated pattern (blue line) and the calculated Bragg positions for a monoclinic unit cell with the space group *C2/m*. The calculated peak positions and intensities are in satisfactory agreement with the experimental data. Table 2 summarizes the results of the structural refinement and the lattice parameters. Note that the presence of a high concentration of stacking faults is a well-known problem in layered vdW compounds and was observed for $Ni_2P_2S_6$ in Ref. [45 -46]. Also stacking faults in the samples manifest by an asymmetry of the *00l* reflections in the pXRD patterns. This might explain the high value of our R-factor from our x-ray analysis.

TABLE II. Structural parameters and residual factors of Rietveld refinement.

| | pXRD |
|---|---|
| *Composition* | $(Mn_{0.7}Ni_{0.3})_2P_2S_6$ |
| *Space group* | C2/m (No. 12) |
| *Wavelength (Å)* | 1.540560 |
| *2θ range (°)* | 10 – 100 |
| *Step Size (°)* | 0.015 |
| *Temperature (K)* | 293 |
| *a (Å)* | 6.009 (7) |
| *b (Å)* | 10.405 (3) |
| *c (Å)* | 6.779 (3) |
| *β (°)* | 107.45 (6) |
| *V (Å³)* | 404.40 (6) |
| *Goodness-Of-Fit* | 2.06 |
| *Bragg R-factor* | 7.61 |
| *RF-factor* | 22.23 |

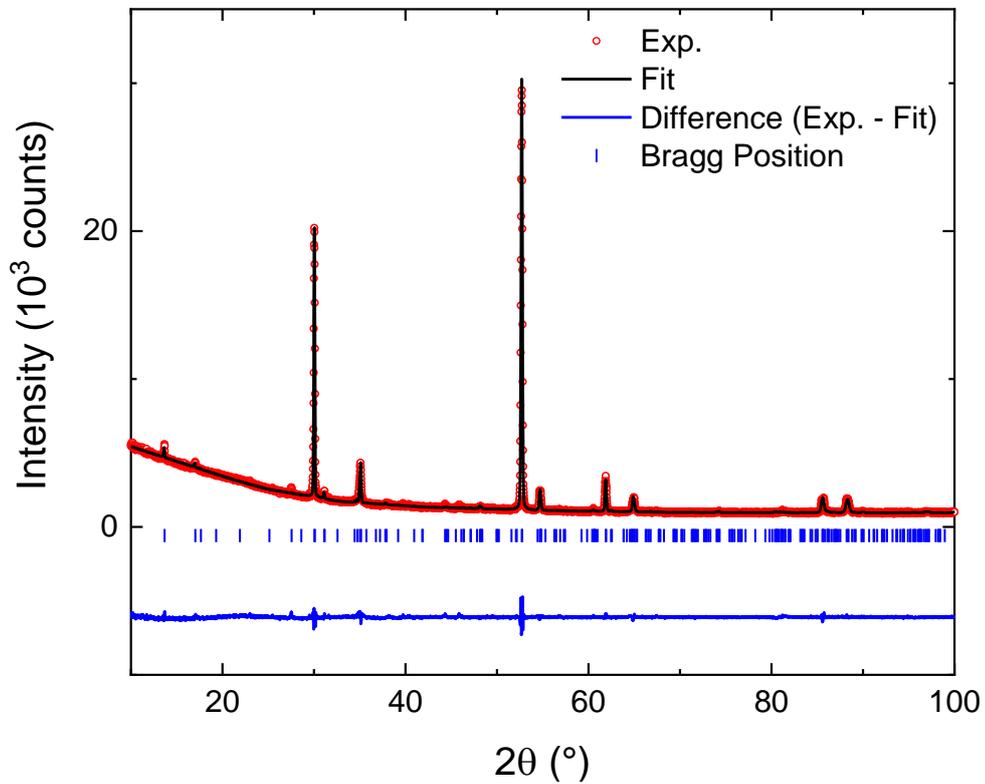

Figure 6. Rietveld analysis of $(Mn_{0.7}Ni_{0.3})_2P_2S_6$.

## 4. Magnetization and transport
   Magnetic properties:

Figure 7. (a-d) show the magnetization measurements as a function of temperature for the entire series of $(Mn_{1-x}Ni_x)_2P_2S_6$ single crystals with a magnetic field of $\mu_0H = 1$T applied parallel and perpendicular to the *ab* plane. The overall behavior of *M/H* is very similar for all compositions with two prominent features in the magnetization curve, i.e., a broad maximum at elevated temperatures, $T_{max}$, related to (low-dimensional) magnetic spin correlations in these quasi-two dimensional van der Waals materials [35-37], together with an inflection point at somewhat lower temperature, $T_N$, signaling the phase transition into a long-range ordered antiferromagnetic state. The Néel temperature $T_N$ is defined from the sharp peak in the derivative $d\chi/dT$ for *H* along the crystallographic *c* axis. Overall, the doping dependence of $T_N$ is non-monotonous (see also Fig. 8): While the end member $Mn_2P_2S_6$ has $T_N \sim 77$ K, with increasing $Ni^{2+}$ content the temperature of the antiferromagnetic phase transition first decreases up to $x = 0.5$, and then increases up to $T_N = 155$ K for $Ni_2P_2S_6$ ($x = 1$).

The single-crystalline nature of our samples allows to further comment on the orientation dependence of the normalized magnetization with respect to the alignment in a magnetic field, *viz.*, the magnetic anisotropy and its evolution as a function of Ni substitution. Below $T_N$, for $Mn_2P_2S_6$ we find $M/H_{\parallel c} < M/H_{\parallel ab}$ with magnetic moments oriented along the *c* direction in agreement with the literature [35]. Upon Ni substitution this anisotropy is gradually suppressed up to about $x \sim 0.7$ before it gets reversed for even higher substitution levels up to $x = 1$ (i.e., for $Ni_2P_2S_6$ the ordered magnetic moments are oriented in the *ab* plane). A clear anisotropic behavior of $T_{max}$ is observed for all intermediate compounds $(Mn_{1-x}Ni_x)_2P_2S_6$ ($x_{Ni} = 0.3, 0.5, 0.7$). It begins to appear already for $x = 0.3$ and reaches the maximum anisotropy at $x = 0.7$ then decreases again.

The observed shift of $T_N$ to lower temperature compared to the parent compound is similar to the behavior of bimetallic $(Mn_{1-x}M_x)_2P_2S_6$ substituted with diamagnetic ions $Mg^{2+}$ or $Zn^{2+}$. In that series, however, for $x > 0.3$ a percolation threshold was observed in Ref. [46-49]. In our $(Fe_{1-x}Ni_x)_2P_2S_6$ substitution series [39], we also observed a gradual evolution of the magnetic transition temperature up to $x = 1$. Note that while $T_N$ shifts in both series, $(Mn_{1-x}Ni_x)_2P_2S_6$ and $(Fe_{1-x}Ni_x)_2P_2S_6$, a minimum in the $T_N(x)$ dependence was only observed for $(Mn_{1-x}Ni_x)_2P_2S_6$. Also, the magnetic anisotropy as well as the broad maximum at $T_{max}$ show two distinctly different behaviors for $(Fe_{1-x}Ni_x)_2P_2S_6$, one up to $x_{Ni} = 0.9$ which is similar to the magnetic anisotropy of $Fe_2P_2S_6$ and another one for $Ni_2P_2S_6$, whereas a gradual but non-monotonous behavior is observed for $T_{max}(x)$ in our $(Mn_{1-x}Ni_x)_2P_2S_6$ series.

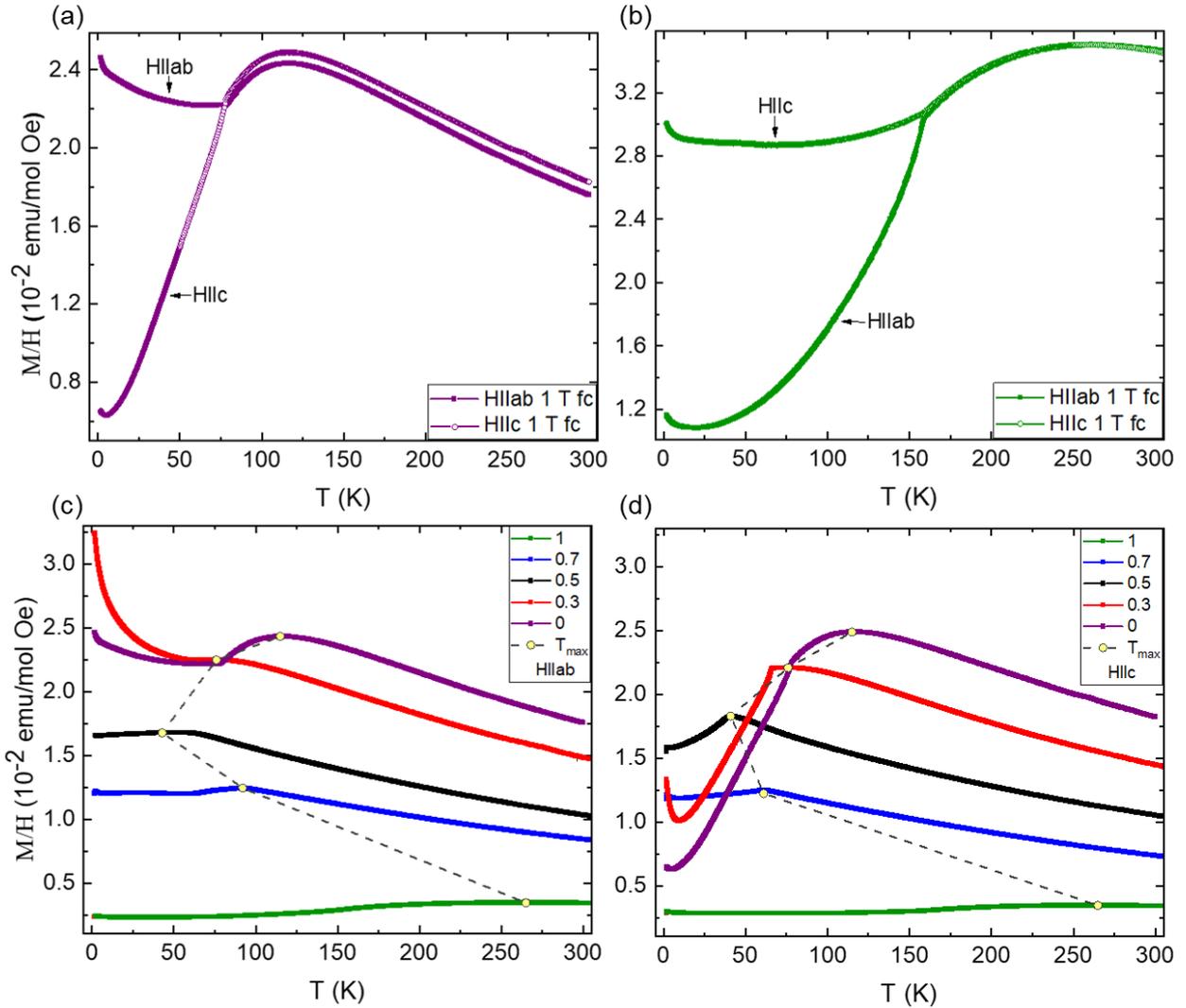

Figure 7. Magnetization as function of temperature for a field of 1 T (a) for $Mn_2P_2S_6$, (b) for $Ni_2P_2S_6$ together with an overview of the magnetization as function of temperature for the full series $(Mn_{1-x}Ni_x)_2P_2S_6$ measured in a field of 1 T applied (c) $H||ab$ and (d) $H||c$

Note that the parent compound $Ni_2P_2S_6$ has a broad maximum well above $T_N$, which gets more pronounced for Mn substitution and becomes the dominant feature for $Mn_2P_2S_6$. The shape of the maximum seems to be determined by the difference between $T_{max}$ and the inflection point. The smaller the difference, the sharper the maximum becomes. $Ni_2P_2S_6$ shows the highest $T_{max}$ as well as the highest $T_N$, whereas $(Mn_{0.5}Ni_{0.5})_2P_2S_6$ shows the lowest $T_{max}$ and $T_N$ and which appear very close to each other in contrast to both parent compounds.

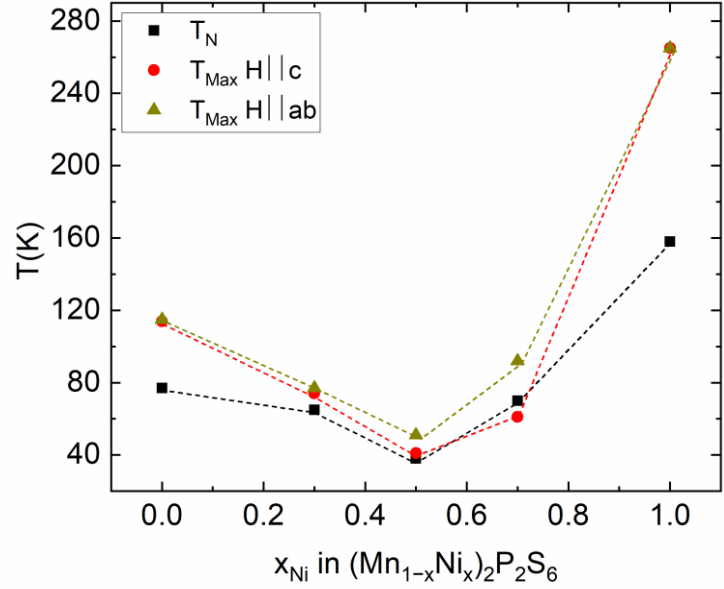

Figure. 8. Evolution of $T_N$ and $T_{max}$ as a function Ni substitution for $(Mn_{1-x}Ni_x)_2P_2S_6$ single crystals.

Figure 9 (a-e) shows the magnetization as a function of the applied field for $(Mn_{1-x}Ni_x)_2P_2S_6$ at $T = 1.8$ K. The field was applied parallel and perpendicular to the *ab* plane. A linear dependence of *M* vs *H* is observed for fields in the *ab* plane for all compositions, except for $x = 1$. The largest difference between the curves for $H\|ab$ and $H\|c$ is observed for higher nickel contents $x = 1$ and 0.7 (Fig. 9 (d-e)). For $x = 0.5$ the curves overlap and then, clear spin-flop transitions are observed at $\mu_0 H_c = 1.9$ T and 3.7 T for $x = 0.3$ and 0, respectively, for magnetic fields applied parallel to the *c* axis (see Fig. 9 (a-b)). The spin-flop is also seen for $x = 0.5$, but it is very weak and broadened. Figure 9 (f) shows the difference $\Delta M = M(H_{\|ab}) - M(H_{\|c})$ as a function of the magnetic field at T = 1.8 K. $(Mn_{1-x}Ni_x)_2P_2S_6$ samples with $x < 0.5$ show a positive pronounced $\Delta M$ in the positive field direction in line with an easy-axis anisotropy. In samples with $x > 0.5$ this behavior is inversed showing an easy-plane anisotropy. The change of the anisotropy is mainly determined via the interplay between the single-ion anisotropy and the crystal field effect (CEF) schemes together with the effects of the spin orbit coupling (SOC) [35]. While the single-ion anisotropy and the trigonal distortion are small/negligible for Mn, they are positive (D > 0) and large for Ni. Such a compositional-dependent switch of the magnetic anisotropy has been observed also in other van der Waals systems such as mixed halides [50]. The observed compositional dependence of the switch of magnetic anisotropy is very appealing for possible applications in the future.

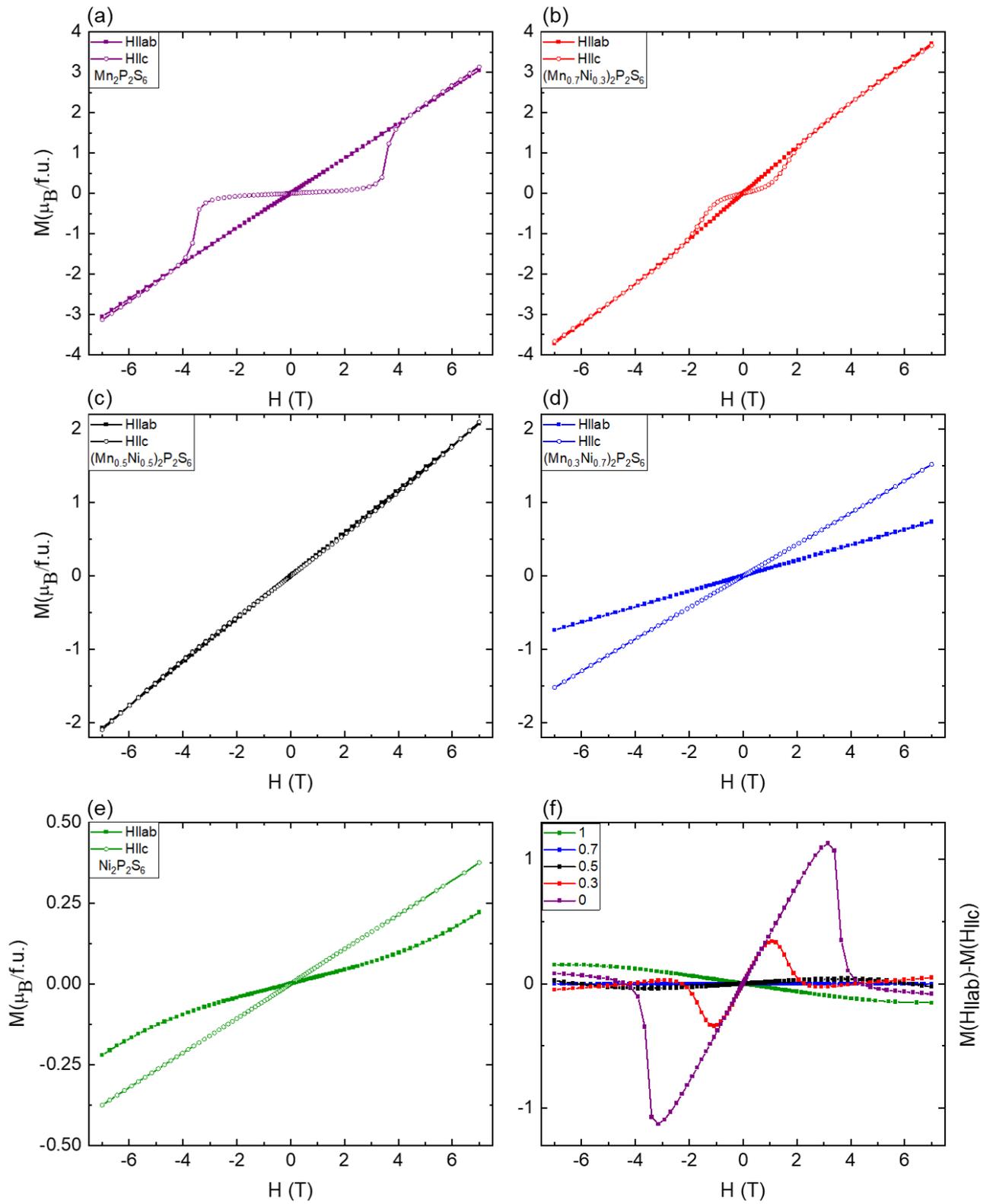

Figure 9. (a-e) Field dependent magnetization at 1.8 K for the full series $(Mn_{1-x}Ni_x)_2P_2S_6$ for a magnetic field applied along $H\|ab$ and $H\|c$. (f) The difference $\Delta M = M(H_{\|ab}) - M(H_{\|c})$ is plotted as a function of magnetic field for all samples.

**Transport properties:**

Electrical resistivity, Hall and magnetotransport measurements were performed on single crystals of $Mn_2P_2S_6$. Figure 10 shows the temperature-dependent resistance $R(T)$ of our $Mn_2P_2S_6$ single crystal at various pressures up to 59.0 GPa. As shown in Fig. 10a, $Mn_2P_2S_6$ displays an insulating behavior at 18.8 GPa, similar to the early reports from Wang et al. [51]. With increasing pressure up to 22.2 GPa, the resistance in the whole temperature range decreases remarkably and a metal-insulator transition appears at $T_{MIT}$~250 K for 22.2 GPa (see Fig. 10b). One can see that $T_{MIT}$ initially decreases up to 25 GPa and then increases in the pressure range 25.0-30.9 GPa. Notably, the sample exhibits a metallic behavior in the whole temperature range up to room temperature at 34.1 GPa. As shown in Fig. 10c, the metallic behavior maintains with increasing pressure up to 59.0 GPa and no traces of superconductivity are detected down to 1.8 K.

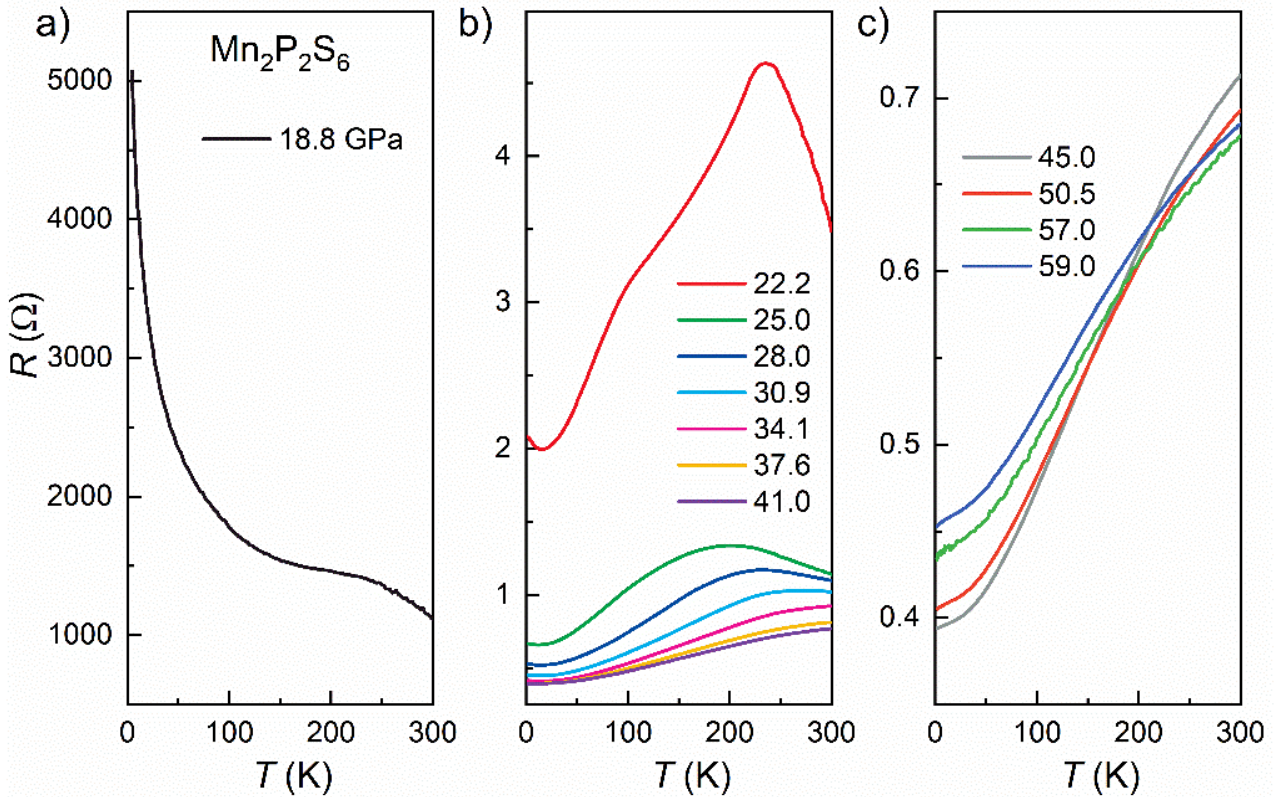

Figure 10. Evolution of the resistivity for $Mn_2P_2S_6$ as function of temperature under different applied hydrostatic pressure.

High-pressure Hall resistance and magnetoresistance measurements were further performed to detect the evolution of charge carriers in pressurized $Mn_2P_2S_6$. Figure 11a shows the selective Hall resistivity curves $R_{xy}(B)$ measured at 10 K under pressure, where the magnetic field is applied perpendicular to the *ab*-plane. Above 18.8 GPa, the $R_{xy}(B)$ curves exhibit a quasi-linear feature with positive slope, indicating hole-dominated charge carriers. According to the single band model, the Hall coefficient $R_H$, extracted from the slope of $R_{xy}(B)$ around zero field, decreases initially with increasing pressure and then increases above 30.9 GPa (see Fig. 11b). Figure 10c displays the magnetoresistance $MR = [(\rho(B) - \rho(0)]/\rho(0) \times 100\%$ at 10 K at various pressures. Note that the *MR* is very small with negative values. Similar to the case of the Hall coefficient, the absolute values of the specific MR at 5 T and 10 K initially decreases and then increases above 30.9 GPa.

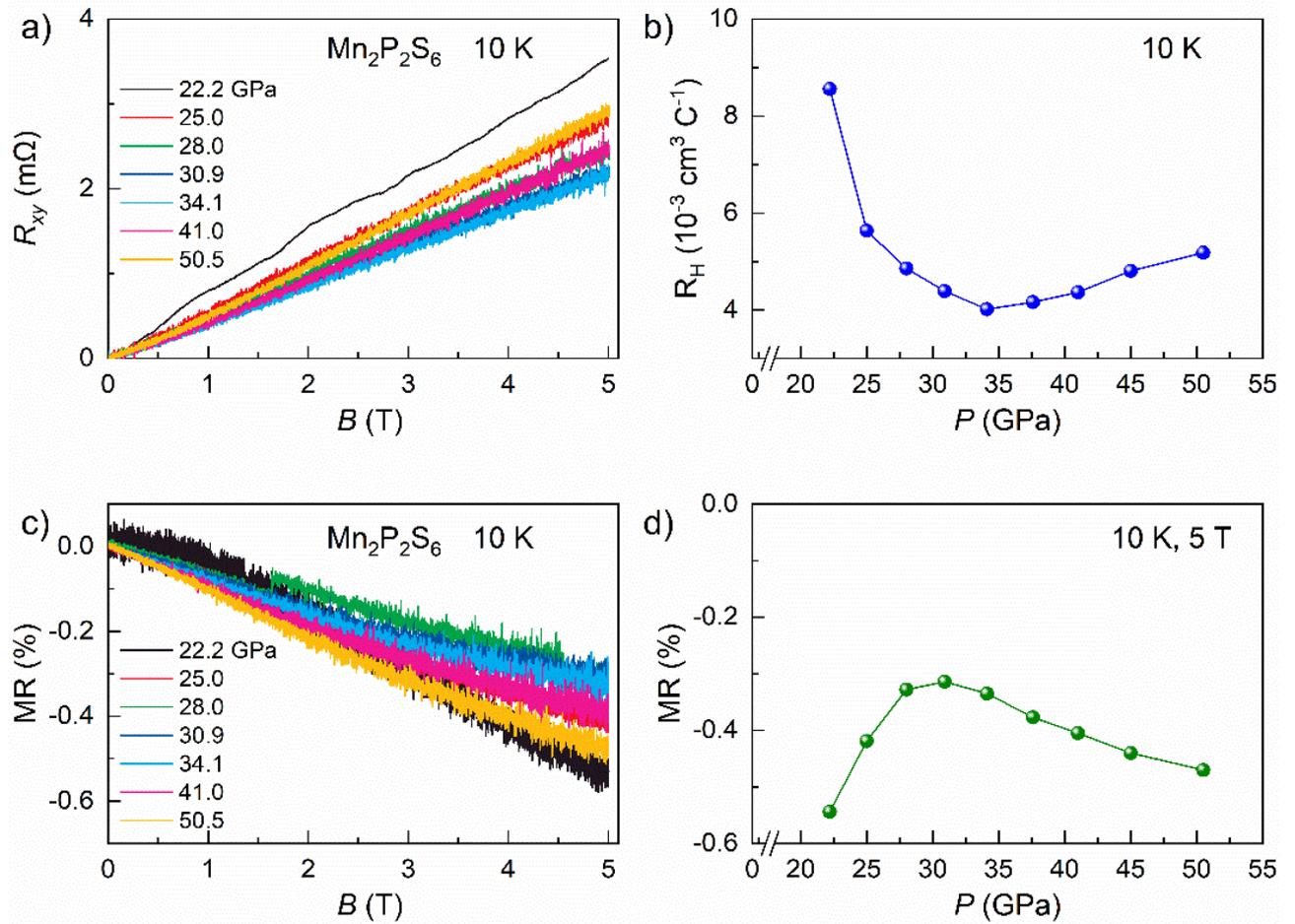

Figure 11. Hall resistivity and magnetoresistance of Mn$_2$P$_2$S$_6$ as function of applied magnetic field under different applied hydrostatic pressures.

# 5. Conclusions

In summary, we have successfully grown single crystals of $(Mn_{1-x}Ni_x)_2P_2S_6$ via the chemical vapor transport technique using Iodine as a transport agent. The optimized crystal growth conditions yielded a series of $(Mn_{1-x}Ni_x)_2P_2S_6$ with (x=0, 0.3, 0.5, 0.7, 1) single crystals up to 4 mm × 3 mm × 200 μm. As-grown crystals obtained by this method were characterized by means of SEM\EDX for compositional analysis and powder x-ray measurements for the structural analysis. Compositional analysis confirms the chemical homogeneity for both parent compounds as well as for the substituted series. The structural characterization shows that all crystals crystallize in the monoclinic space group *C2/m* (No. 12). The magnetic measurements as a function of temperature were performed for applied fields parallel and perpendicular to the crystallographic *ab* plane. All as-grown single crystals show the presence of long-range antiferromagnetic order. The end member $Ni_2P_2S_6$ has the highest $T_N$ ~155K and which non-monotonously shifts to lower temperature as a function of Mn substitution. The shift in the long-range magnetic order and the gradual change in the anisotropy shows some similarity to the sister compounds like $(Fe_{1-x}Ni_x)_2P_2S_6$ and yet there are some significant differences, such as the lowest $T_N$ and $T_{max}$ for the $(Mn_{0.5}Ni_{0.5})_2P_2S_6$ composition. The magnetic anisotropy switches from out-of-plane to in-plane as a function of composition in the $(Mn_{1-x}Ni_x)_2P_2S_6$ series. Our electrical transport measurements on the end member $Mn_2P_2S_6$ clearly indicates an insulator-to-metal transition at a critical pressure of ~ 22 GPa. For pressures higher than 40 GPa, $Mn_2P_2S_6$ becomes metallic at room temperature. The magnetoresistance measured at 10K shows a negative magnetoresistance. The successful growth of high-quality single crystals of our series $(Mn_{1-x}Ni_x)_2P_2S_6$ opens an opportunity for further anisotropic investigations in the future.

### Acknowledgements

S.A. acknowledges the support of Deutsche Forschungsgemeinschaft (DFG) through Grant No. AS 523/4–1. A.U.B.W. and B.B. acknowledge financial support from the DFG through SFB 1143 (project-id 247310070) and the Würzburg-Dresden Cluster of Excellence on Complexity and Topology in Quantum Matter – ct.qmat (EXC 2147, project-id 390858490).

**\* Corresponding author:** s.aswartham@ifw-dresden.de